\documentclass[aps,pre,preprint,groupaddress,floats,showpacs]{revtex4-1}
\usepackage{amssymb}
\usepackage{graphicx}
\usepackage{dcolumn}
\usepackage{bm}
\usepackage{txfonts}
\begin{document}
\preprint{APS/Version 1}
\title[Information directionality via transcripts]{Information directionality in coupled time series using transcripts}
%\title[]{Dimensional reduction by connected transcripts}
\author{Roberto Monetti}
\email{monetti@mpe.mpg.de}
\affiliation{Max-Planck-Institut f\"ur extraterrestrische Physik,
Giessenbachstr. 1, 85748 Garching, Germany}
\author{Wolfram Bunk}
\affiliation{Max-Planck-Institut f\"ur extraterrestrische Physik,
Giessenbachstr. 1, 85748 Garching, Germany}
\author{Thomas Aschenbrenner}
\affiliation{Max-Planck-Institut f\"ur extraterrestrische Physik,
Giessenbachstr. 1, 85748 Garching, Germany}
\author{Stephan Springer}
\affiliation{Klinik Hochried, Zentrum f\"ur Kinder, Jugendliche und Familien, Hochried 1 - 12, 82418 Murnau, Germany}
\author{Jos\'e M. Amig\'o}
\affiliation{Centro de Investigaci\'on Operativa, Universidad Miguel
Hernandez, Avda. de la Universidad s/n, 03202 Elche, Spain}
%\footnote{Present address:Department of Physics, University of Bristol, Tyndalls Park Road, Bristol BS8 1TS, UK.} 
\date{\today}
\begin{abstract}
In ordinal symbolic dynamics, transcripts describe the algebraic relationship between ordinal patterns. Using the concept of transcript, we exploit the mathematical structure of the group of permutations to derive properties and relations among information measures of the symbolic representations of time series. These theoretical results are then applied for the assessment of coupling directionality in dynamical systems, where suitable coupling directionality measures are introduced depending only on transcripts. These novel measures estimate information flow in lower space dimension and reduce to well-established coupling directionality quantifiers when some general conditions are satisfied. Furthermore, by generalizing the definition of transcript to ordinal patterns of different lengths, several of the commonly used  information directionality measures can be encompassed within the same framework.
\end{abstract}
\maketitle

\section{Introduction}
The study of dynamical behavior in interacting complex systems is relevant in different fields of science \cite{Pikovskibook,Lehnertzbook}. Developments in the area of non-linear dynamics and the use of information theoretic approaches have greatly contributed to the understanding of ubiquitous phenomena like synchronization \cite{Glass2001} and collective behavior in spatially extended systems \cite{Tononi1998,Strogatz2005}. Great attention has recently been paid to the study of causality and the assessment of coupling directionality in dynamical systems \cite{Sugihara2012,Nolte2008,Vejmelka2008,Frenzel2007}. Granger causality \cite{Granger1969} was probably the first method which introduced the notion of predictability to detect interaction asymmetry in linear models. Using the concept of Granger causality other directionality measures were proposed to account for non-linear interactions in dynamical systems \cite{Schiff1996,Faes2008}. Apart from the traditional methods based on information theoretic concepts \cite{Schreiber2000,Palus2003,Frenzel2007,Vejmelka2008}, other authors have suggested the use of non-linear state space reconstruction \cite{Sugihara2012} and the phase-slope of cross spectra \cite{Nolte2008}.
The characterization and detection of information flow has also been investigated from the viewpoint of ordinal symbolic dynamics \cite{Amigobook}. Several approaches have been proposed suggesting advantages of the use of ordinal symbolic dynamics like computational efficiency and robustness against noise \cite{Staniek2008,Kantz2008,Pompe2011,Papana2011}.

Ordinal time series analysis is a particular form of symbolic analysis whose
\textquotedblleft symbols\textquotedblright\ are ordinal patterns of a given
length $L\geq 2$. This concept was introduced by C. Bandt and B. Pompe in
their seminal paper \cite{Bandt2002}, in which they also introduced
permutation entropy as a complexity measure of time series. Since then,
ordinal time series analysis has found a number of interesting applications
in biomedical sciences, physics, engineering, finance, statistics, etc.
One important aspect of this new tool in data analysis is the fact that the
ordinal patterns of length $L$, which can be identified with permutations of 
$L$ objects, have a well-known mathematical structure. Indeed, permutations
build a (non-commutative) multiplicative group called the symmetric group of
order $L$. 
The mathematical structure of the symmetric group is exploited by the concept of
transcript.
Transcripts were introduced in \cite{Monetti2009} and applied for characterizing the
synchronization behavior of two coupled, chaotic oscillators. In this work
we will present a further application, this time to the characterization of
the coupling directionality between time series.
\section{Theoretical setting \label{sec:theo}}
Let $(x_{n})_{n\in \mathbb{N}_{0}}$ be a sequence whose
elements $x_{n}$ belong to a set  endowed with a total ordering $\leq $. The $L$-block $x_{n}^{n+(L-1) \text{T}}=x_{n},x_{n+\text{T}},...,x_{n+(L-1)\text{T}}$ can be associated to the \textit{ordinal} $L$-\textit{pattern} $\pi =\left\langle \pi _{0},...,\pi _{(L-1)}\right\rangle$ as follows, 
\[
x_{n+\pi _{0}\text{T} }<x_{n+\pi _{1}\text{T}}<...<x_{n+\pi _{(L-1)}\text{T}},
\]%
where in case $x_{i}=x_{j}$, we agree to set $x_{i}<x_{j}$ if, say, $i<j$. Here, $\text{T} \geq 1$ is a time delay used for the construction of ordinal patterns.
Therefore, an ordinal $L$-pattern (or ordinal patterns
of length $L$) is the permutation of the integer numbers $0$, 
$1$,..., $L-1$ indicating the rank ordering (according to their size) of the elements 
$x_{n},x_{n+\text{T}},...,x_{n+(L-1)\text{T}}$, where $n$ is arbitrary,  $\text{T}\geq 1$, and $L\geq 2$.
Specifically, $\pi =\left\langle \pi _{0},...,\pi _{(L-1)}\right\rangle $ may
be identified with the permutation $i\mapsto \pi _{i}$, $0\leq i\leq (L-1)$.

The set of ordinal $L$-patterns forms a finite non-Abelian group of order $L!$ (the so-called \textit{symmetric group} $\mathcal{S}_{L}$), when equipped with the product of permutations defined as
\begin{equation}
\pi \circ \sigma = \left\langle  \sigma_{\pi_0}, \sigma_{\pi_1}, \ldots , \sigma_{\pi_{L-1}} \right\rangle,
\label{prod}
\end{equation}
with the inverse element being given by
\[
\pi ^{-1}=o(\pi _{0},...,\pi _{L-1}), 
\]%
and the unity by the identity permutation, 
\[
id=\left\langle 0,1,...,L-1\right\rangle. 
\]
Here, $o$ denotes the sorting operation. For example $o(2,0,1)=\left\langle1, 2, 0 \right\rangle$.

The algebraic structure of $\mathcal{S}_{L}$ is exploited by the concept of
transcripts. 
In fact, being  $\mathcal{S}_{L}$ a group, 
given $\alpha ,\beta
\in \mathcal{S}_{L}$, there always exists a \emph{unique} $\tau =\tau_{\alpha \beta}
\in \mathcal{S}_{L}$, called \textit{transcript } from the \textit{source
pattern} $\alpha $ to the \textit{target pattern} $\beta $, such that 
\begin{equation}
\tau \circ \alpha =\beta ,  \label{2_transcription}
\end{equation}%
where $\tau \circ \alpha =\left\langle \alpha _{\tau _{0}},\alpha _{\tau
_{1}},...,\alpha _{\tau _{L-1}}\right\rangle $ (see Eq.~(\ref{prod}))$.$ It follows
that $\tau $ is a transcript from $\alpha $ to $\beta $ if and only if $\tau
^{-1}$ is a transcript from $\beta $ to $\alpha $. As usual, we will write
hereafter the product of $\alpha $ and $\beta $ just as $\alpha \beta$, unless otherwise convenient.
As the source pattern $\alpha $ and the target pattern $\beta $ vary over $\mathcal{S}_{L}$, their transcript varies according to $\tau_{\alpha \beta}
=\beta \circ \alpha ^{-1}$. Note that different pairs $(\alpha ,\beta )$ can share the same transcript. 
More generally, given $\tau \in 
\mathcal{S}_{L}$, there exist $L!$ pairs $(\alpha ,\beta )\in \mathcal{S}%
_{L}\times \mathcal{S}_{L}$ such that $\tau $ is the transcript from $\alpha 
$ to $\beta $. 
Two trivial properties of the transcripts are
\begin{equation}
\tau_{\beta,\alpha}=(\tau_{\alpha,\beta})^{-1} \label{trv1}%
\end{equation}
and
\begin{equation}
\tau_{\beta,\gamma}\tau_{\alpha,\beta}=\gamma\beta^{-1}\beta\alpha^{-1}%
=\beta\gamma^{-1}=\tau_{\alpha,\gamma}. \label{trv2}%
\end{equation}
which implies the transitivity of the transcription operation.
For more properties of the transcripts, see \cite{Monetti2009,Amigo2012}.

Consider two stationary time series $\{x_{t}\}$, $\{y_{t}\}$. In turn, they
provide two sequences of $L$-ordinal patterns, $\{\alpha
_{k}\}$ and $\{\beta _{k}\}$, respectively. 
Let $p_{L}^{1}(\alpha )$ ($p_{L}^{2}(\beta )$) be the probability for the source (target) $L$-pattern $\alpha$ ($\beta $) to occur in $\{\alpha_{k}\}$ ($\{\beta _{k}\}$), and $p_{L}^{J}(\alpha ,\beta )$ the joint probability.
Then, the probability function of the transcripts, $p_{L}^{T}(\tau )$, $\tau \in \mathcal{S}_{L}$,
can be written as
\[
p_{L}^{T}(\tau )=\sum_{(\alpha ,\beta ): \beta \alpha^{-1}=\tau}p_{L}^{J}(\alpha ,\beta ),
\]
Thus, the entropy of the joint probability function $p_{L}^{J}$ and the entropy of the corresponding transcript probability function $p_{L}^{T}$ are defined as
\[
H(\alpha,\beta)=-\sum\limits_{\alpha ,\beta \in \mathcal{S}_{L}}p_{L}^{J}(%
\alpha ,\beta )\log p_{L}^{J}(\alpha ,\beta ),
\]%

\noindent and

\[
H(\tau)=-\sum\limits_{\tau \in \mathcal{S}_{L}}p_{L}^{T}(\tau )\log
p_{L}^{T}(\tau ),
\]%
respectively, where we have used $H(\alpha,\beta)=H(p_{L}^{J})$ and $H(\tau)=H(p_{L}^{T})$ for notational convenience.

The definition of transcripts given by Eq.~(\ref{2_transcription}), provides the algebraic relationship between source and target ordinal patterns. It follows that, given the triple $(\alpha, \beta,\tau)$, the knowledge of any pair of symbols, i.e. $(\alpha, \beta)$, $(\alpha, \tau)$, or  $(\beta,\tau)$, univocally determines the remaining symbol. This important property implies 
\begin{equation}
H(\alpha,\beta)=H(\alpha,\tau)=H(\beta,\tau).
\label{igual}
\end{equation}
More general, given the random variables $\alpha^{n}$, $1\leq n\leq N$, with outcomes in
$\mathcal{S}_{L}$, then
\begin{eqnarray}
H(...,\alpha^{n},\alpha^{n+1},...)  &  =H(...,\alpha^{n},\tau_{\alpha
^{n},\alpha^{n+1}},...)=H(...,\alpha^{n},\tau_{\alpha^{n+1},\alpha^{n}},...)\label{trv3a}\\
&  =H(...,\tau_{\alpha^{n},\alpha^{n+1}},\alpha^{n+1},...)=H(...,\tau
_{\alpha^{n+1},\alpha^{n}},\alpha^{n+1},...) \label{trv3b}
\end{eqnarray}
because any of the random variable pairs explicitly shown in (\ref{trv3a})-(\ref{trv3b}) can be determined from any other variable pair.

The concept of coupling complexity was first introduced in \cite{Amigo2012} along with two complexity indices for its quantification. Coupling complexity refers to the relationship among dynamical system components; in general, it differs from the complexity of the individual components or from their sum. Here, we consider only one of two coupling complexity indices proposed,  namely
\begin{equation}
C(\alpha , \beta)=\min \{H(\alpha),H(\beta)\}  -(H(\alpha,\beta)-H(\tau) ).
\label{theorem}
\end{equation}
By means of Eq.~(\ref{igual}), $C(\alpha,\beta)$ can be written as
\begin{equation}
\label{ccmi}
C(\alpha , \beta) = \min \{I( \alpha,\tau),I(\beta,\tau)\},
\end{equation}
where $I$ denotes mutual information. As mutual information is a positive definite quantity, we demonstrated here again that $C(\alpha , \beta)  \geq 0$. 
The complexity index $C(\alpha , \beta) $ can also be written as
\begin{equation}
C(\alpha , \beta)=H(\tau)-\max\{H( \alpha \mid \beta),H( \beta \mid \alpha)\},
\label{c1cond}
\end{equation}
where $H( \alpha \mid \beta )$ is a conditional entropy. Since $C(\alpha , \beta) \geq 0$, Eq.~(\ref{c1cond}) implies $H(\tau) \geq \max\{H( \alpha \mid \beta),H( \beta \mid \alpha)\}$. 
The complexity can be generalized to multivariate time series analysis by means of the following expression
\begin{equation}
C(\alpha^1, \alpha^2, \ldots , \alpha^m)=\min \{H(\alpha^1),H(\alpha^2), \ldots ,  H(\alpha^m)\}+ H(\tau_{12},\tau_{23}, \ldots , \tau_{(m-1)m})-H(\alpha^1, \alpha^2, \ldots , \alpha^m) 
\label{c1hdd} 
\end{equation}
\begin{equation}
C(\alpha^1, \alpha^2, \ldots , \alpha^m)=\min_{1 \leq i \leq m} I(\alpha^i; \tau_{12},\tau_{23}, \ldots , \tau_{(m-1)m}),
\label{c1hd}
\end{equation}
where $\alpha^n$ denotes the symbolic representation of the $n^{th}$ time series and $\tau_{(n-1)n}$ are the transcripts connecting symbolic representations $\alpha^{n-1}$ and $\alpha^{n}$. A proof of (\ref{c1hd}) is presented in \cite{Monetti2013}.
Similarly to the bivariate case, the generalized coupling complexity is invariant under the interchange of the $\alpha^{n}$'s. 
For instance, consider three symbolic representations $\{\gamma_i \}$, $\{\beta_i \}$, and $\{\alpha_i \}$, and all possible transcripts $\{(\tau_{\gamma,\beta})_i\}$,  $\{(\tau_{\gamma,\alpha})_i\}$, and $\{(\tau_{\beta,\alpha})_i\}$. Since given two of the three transcripts $\tau_{\gamma,\beta}$,  $\tau_{\gamma,\alpha}$, and $\tau_{\beta,\alpha}$ the third one can be determined via (\ref{trv1}) and (\ref{trv2}), it follows that
$H(\tau_{\gamma,\beta},  \tau_{\gamma,\alpha}) = H(\tau_{\gamma,\beta},  \tau_{\beta,\alpha})=H(\tau_{\gamma,\alpha},\tau_{\beta,\alpha})$ and therefore the invariance of $C(\alpha,\beta , \gamma)$ (see Eq.~(\ref{c1hd})) under permutation of its arguments. 
For a general proof of this property see \cite{Monetti2013}.
\section{Information directionality}
\subsection{Methods}
The detection of the coupling direction between dynamical systems requires 
asymmetric measures sensitive to the part of information not contained in the joint past of the systems. The conditional mutual information (CMI) is such a quantity, having been already used in several applications \cite{Palus2003a,Palus2003}. We will consider the CMI within the framework of ordinal symbolic dynamics as already proposed in different approaches \cite{Staniek2008,Kantz2008}. 
First, we generate symbolic representations and transcripts for coupled dynamical systems using length $L$ and delay T. 
Let $\{\alpha_i\}$, $\{\beta_i\}$, $\{\gamma_i \}$ be three symbolic representations. The CMI can be written as follows
\begin{equation}
I(\gamma,\beta\mid\alpha)=H(\gamma\mid\alpha)-H(\gamma \mid \beta , \alpha).
\label{cmi}
\end{equation}
For  $\{\gamma_i\}=\{\alpha_{i+\Lambda} \}$, with $\Lambda > 0$, Eq. (\ref{cmi}) becomes a measure of coupling directionality between two dynamical systems, namely the symbolic transfer entropy $T_{X,Y}^S$ introduced in \cite{Staniek2008}. Thus, using the asymmetry of the CMI under the interchange of the time series, one can easily construct indices of information flow, for instance the difference $T_{X,Y}^S - T_{Y,X}^S$. 

Now, we introduce and motivate the use of a new coupling directionality measure based on the mutual information of transcripts defined as follows,
\begin{equation}
I(\tau_{\gamma,\alpha},\tau_{\beta,\alpha})=H(\tau_{\gamma,\alpha})-H(\tau_{\gamma,\alpha} \mid \tau_{\beta,\alpha}).
\label{rr}
\end{equation}
First, note that Eq.~(\ref{rr}) is only a function of transcripts between symbolic representations. Furthermore, it displays the same invariance under the interchange of $\gamma$ and $\beta$ and asymmetry when interchanging the roles played by $\alpha$ and $\beta$ as Eq. (\ref{cmi}). Having in mind that transcripts account for the relationship between symbolic representations, one can discover qualitative similarities between Eqs.~(\ref{cmi}) and (\ref{rr}). In fact, one observes that stronger (weaker) dependence between $\beta$ and $\gamma$, increases (decreases) both informations given by  Eqs.~(\ref{cmi}) and (\ref{rr}).  However, a relevant difference is evident in Eq.~(\ref{rr}), i.e. the estimate of  information flow is calculated in lower dimension. 

Let us assume again that $\{\gamma_i\}=\{\alpha_{i+\Lambda} \}$ and consider the case $\{\beta_i\}$ independent of $\{\alpha_i\}$ and $\{\gamma_i \}$. Clearly, $I(\gamma,\beta\mid\alpha)=0$ in this case. We are going to show next that the same property holds for $I(\tau_{\gamma,\alpha},\tau_{\beta,\alpha})$ under the additional assumption that $\alpha$ (hence $\gamma$) or $\beta$ are uniformly distributed. Indeed, 
using that $C(\gamma,\alpha,\beta) \geq 0$, Eq.~(\ref{rr}) can be bounded as (see (\ref{c1hd}) with $m=3$)
\begin{eqnarray}
I(\tau_{\gamma,\alpha},\tau_{\beta,\alpha}) &\equiv& H(\tau_{\beta,\alpha})+H(\tau_{\gamma,\alpha}) - H(\tau_{\gamma,\alpha},\tau_{\beta,\alpha})  \nonumber \\
&\leq& H(\tau_{\beta,\alpha})+H(\tau_{\gamma,\alpha})+\min\{ H(\gamma), H(\alpha),  H(\beta) \} - H(\gamma, \alpha, \beta). \nonumber
\end{eqnarray}
Here, $H(\gamma)=H(\alpha)$ and $H(\gamma, \alpha, \beta) = H(\beta)+H(\gamma,\alpha)$ since we assumed independence. The latter expression can also be written as $H(\gamma, \alpha, \beta) = H(\alpha,\beta)+H(\gamma,\alpha)-H(\alpha)$.  Therefore,
\begin{equation}
I(\tau_{\gamma\alpha},\tau_{\beta\alpha}) \leq H(\tau_{\beta,\alpha})+H(\tau_{\gamma,\alpha})+\min\{ H(\alpha), H(\beta)  \}- H(\alpha,\beta) -H(\gamma,\alpha)+H(\alpha).
\label{ineqq}
\end{equation}
Using Eq.~(\ref{igual}), $H(\alpha,\beta) = H(\tau_{\beta\alpha},\beta)=H(\tau_{\beta\alpha},\alpha)$ and $H(\gamma,\alpha)=H(\tau_{\gamma\alpha},\alpha)$. Let us assume now that the variable $\beta$ is uniformly distributed. Then, $\min\{ H(\beta), H(\alpha) \}=H(\alpha)$ and $H(\alpha,\beta)=H(\tau_{\beta,\alpha},\alpha)=H(\tau_{\beta,\alpha})+H(\alpha)$, where in the latter expression we used again the independence of $\alpha$ and $\beta$.
Thus, inequality~(\ref{ineqq}) becomes 
\begin{equation}
I(\tau_{\gamma,\alpha},\tau_{\beta,\alpha}) \leq H(\tau_{\gamma,\alpha})+H(\alpha)-H(\tau_{\gamma,\alpha},\alpha).
\label{redunif}
\end{equation}
Similarly, if the variable $\alpha$ is uniformly distributed then $\min\{ H(\beta), H(\alpha) \}=H(\beta)$ and $H(\alpha,\beta)=H(\tau_{\beta,\alpha},\beta)=H(\tau_{\beta,\alpha})+H(\beta)$. Replacing these equations in (\ref{ineqq}), we obtain again Eq.~(\ref{redunif}).
It should be noted that the right hand side of (\ref{redunif}) is independent of the variable $\beta$. As shown below, distributions closer to the uniform distribution can be obtained by a suitable choice of the parameter T. In addition, in case of independence the upper bound in  Eq.~(\ref{redunif}) can be made negligible using a convenient relation between T and $\Lambda$. 

The selection of embedding parameters is a common problem which has been extensively discussed in the field of non-linear systems \cite{Fraser1986}. Directionality measures are not the exception \cite{Pompe2011}. We present in the following an example intended to show the dependency of the directionality measures (\ref{cmi}) and (\ref{rr}) on the parameter T (time delay used to generate the ordinal pattern) for constant $L=4$. 
Consider the following bidirectionally delayed-coupled logistic map $f:[0,1]\rightarrow \lbrack 0,1],\,f(x)=4x(1-x)$ defined by the equations
\begin{eqnarray}
&&x(t)=f(g_{y \rightarrow x} \bmod \, 1),\mbox{with}  \nonumber
\label{clm} \\
&&g_{y \rightarrow x}=k_{1}y(t-\Delta_{1})+(1-k_{1})x(t-1),
\nonumber \\
&&y(t)=f(g_{x\rightarrow y} \bmod \, 1),\mbox{with} \\
&&g_{x\rightarrow y}=k_{2}x(t-\Delta_{2})+(1-k_{2})y(t-1), \nonumber
\end{eqnarray}%
where $\Delta_{1}=5$ and $\Delta_{2}=2$ are the coupling delays, and $k_{1}\in \lbrack 0,1\rbrack$ and $k_{2}\in \lbrack 0,1]$ are the coupling strengths. 
We investigate the coupled logistic map (\ref{clm}) for the coupling parameters $k_2 = 0.2$ and $k_1 \in [0,1]$ as in reference \cite{Pompe2011}. Let  $\{\alpha_i\}$, $\{\beta_i\}$ be the symbolic representations of the time series $\{x_i\}$, $\{y_i\}$, respectively. For every value of $k_1$, we have evaluated the measures defined in Eqs.~(\ref{cmi}) and (\ref{rr}) for several time delays T and time lags $\Lambda \in [-10,10]$. Typically the response of the coupling directionality measures displays a maximum for a certain value $\Lambda = \Lambda_m$. For this system,  $\Lambda_m = 4$ leads to a good description of the information directionality \cite{Pompe2011}.  

\begin{figure}[tbp]
\centering
\includegraphics[width=13.cm,angle=0]{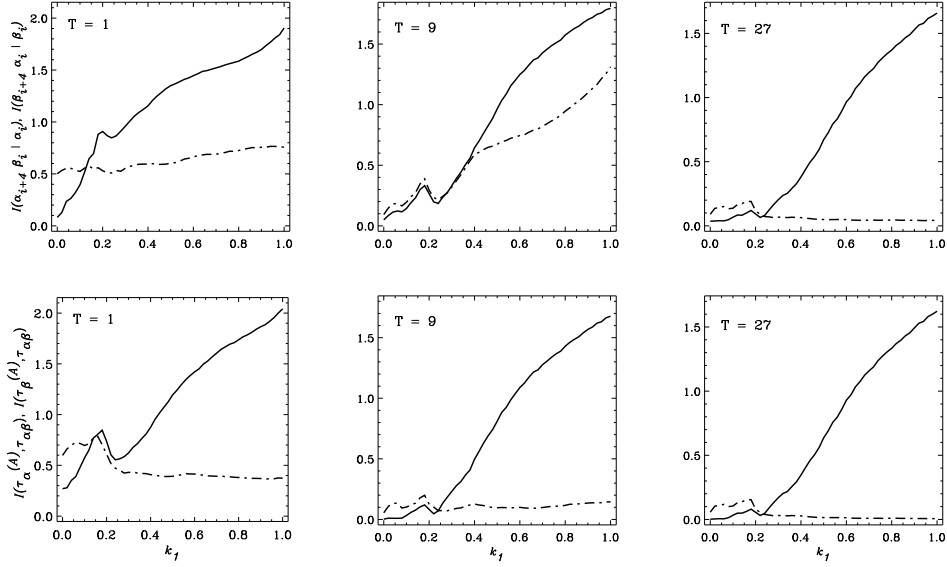}
\caption{Upper row: Conditional mutual informations $I(\alpha_{i+4},\beta_i \mid \alpha_i)$ (solid curve) and $I(\beta_{i+4},\alpha_i \mid \beta_i)$ (dashed curve) versus $k_1$. 
Lower row: The mutual informations $I(\tau_{\alpha}^{(A)} , \tau_{\beta,\alpha})$ (solid curve) and $I(\tau_{\beta}^{(A)} , \tau_{\alpha ,\beta})$ (dashed curve) versus $k_1$, where $\tau_{\alpha}^{(A)} \alpha_i =\alpha_{i+4}$, $\tau_{\beta}^{(A)} \beta_i =\beta_{i+4}$, and $\tau_{\beta,\alpha} \alpha_i =\beta_{i}$. Different panels show the behavior of the coupling directionality measures (Eqs.~(\ref{cmi}) and (\ref{rr})) for different values of T. 
All results were obtained for the coupled logistic map (\ref{clm}) using $L=4$, $\Lambda = 4$ and times series of length $N = 10^5$ data points.
}
\label{fig:1o}
\end{figure}
Figure \ref{fig:1o} shows the behavior of the coupling directionality measures (\ref{cmi}) and (\ref{rr}) versus $k_1$ for different values of the time delay T. In general both measures are able to describe correctly the overall coupling directionality. In fact, we observe that for $k_1 < 0.2$ the direction of information is $x \rightarrow y$, but a crossover to $y \rightarrow x$ is observed when increasing the coupling constant $k_1$, as expected from Eq. (\ref{clm}). Note that the solid (dashed) curves in Fig.~\ref{fig:1o} describe the information flow $y \rightarrow x$ ($x \rightarrow y$), respectively.
However, particular details are observed for different values of the delay time T. Here, $\tau_{\alpha}^{(A)}$ and $\tau_{\beta}^{(A)}$ denote the transcripts between ordinal patterns of the same symbolic representation at different times, as explained in the caption of Fig.~\ref{fig:1o}. For
$\text{T}=1$ and $k_1 =0$, $I(\tau_{\alpha}^{(A)} , \tau_{\beta,\alpha})$ (solid curve) displays a bias to positive values, while $I(\alpha_{i+4},\beta_i \mid \alpha_i) \sim 0$ (solid curve), as expected. For increasing $k_1$, both $I(\tau_{\alpha}^{(A)} , \tau_{\beta,\alpha})$ and $I(\alpha_{i+4},\beta_i \mid \alpha_i)$ increase rather monotonically, except around the value $k_1 \sim 0.20$. For $k_1 \lesssim 0.20$ both $I(\beta_{i+4},\alpha_i \mid \beta_i)$ and $I(\tau_{\beta}^{(A)} , \tau_{\alpha,\beta})$ (dashed curves) indicate the right direction of information flow, but for increasing $k_1$, $I(\beta_{i+4},\alpha_i \mid \beta_i)$ displays a strong unexpected increasing trend. In contrast, $I(\tau_{\beta}^{(A)} , \tau_{\alpha,\beta})$ (dashed curve) remains rather constant. 

For $\text{T}=9$, $I(\alpha_{i+4},\beta_i \mid \alpha_i)$ and $I(\tau_{\alpha}^{(A)} , \tau_{\beta,\alpha})$ describe correctly the coupling in the direction  $y \rightarrow x$. It should be noted that for this value of the delay time, $I(\tau_{\alpha}^{(A)} , \tau_{\beta,\alpha}) \sim 0$ for $k_1 = 0$. 
However,  $I(\beta_{i+4},\alpha_i \mid \beta_i)$ (dashed curve) provides a poor description of the coupling directionality, displaying an even stronger trend than that observed for $\text{T}=1$. On the other hand, $I(\tau_{\beta}^{(A)} , \tau_{\alpha,\beta})$ provides a better description, but still displaying a weak increasing trend for larger $k_1$. For $\text{T}=27$, both  measures provide the same description of the coupling directionality in the system and can rather be distinguished by eye inspection. In fact, we demonstrate below that under certain conditions both coupling directionality measures are identical.

Let us assume that $\min\{H(\alpha),H(\beta)\}=H(\alpha)$ and that the following relation 
\begin{equation}
C(\alpha,\beta,\gamma)=C(\alpha,\gamma)+C(\alpha,\beta),
\label{eqsepa}
\end{equation}
holds for a particular choice of the embedding parameters $L$ and T. For $\{ \gamma_i\} =\{ \alpha_{i+\Lambda}\}$,  Eq.~(\ref{eqsepa}) indicates that the coupling complexity of the three symbolic representations can be expressed as the sum of two terms, namely an "auto"-coupling complexity $C(\alpha,\gamma)$ and a "cross"-coupling complexity $C(\alpha,\beta)$.
Using Eq. (\ref{eqsepa}) one obtains 
\begin{equation}
H(\alpha,\gamma) - H(\alpha) -H(\alpha,\beta,\gamma)+H(\alpha,\beta)=H(\tau_{\gamma,\alpha})+H(\tau_{\beta,\alpha})-H(\tau_{\gamma,\alpha},\tau_{\beta,\alpha}),
\end{equation}
which immediately implies the equality of Eqs.~(\ref{cmi}) and (\ref{rr}).
Thus, we have demonstrated that the CMI estimator can be reduced to the mutual information of transcripts when Eq.~(\ref{eqsepa}) is fulfilled.
The dimensional reduction can be very
important in time series analysis because the number of $N$ joint symbols
grows exponentially with $N$, while the length of real-world time series is
finite. Therefore, the use of expressions similar to Eq. (\ref{rr}) may in some cases prevent from
undersampling and, in any case, it improves the statistical significance of
the estimations.

Another interesting condition which deserves special attention is $C(\gamma,\alpha, \beta) = 0$. 
This particular case is relevant for
a wide range of systems, where a low complexity can be achieved by generating symbolic representations using a suitable time delay T. Typically, the dependence of $C$ on T is such that  $C(\text{T})$ decreases when T grows.
This condition can be compared to that of maximizing the {\it sorting entropy} \cite{Bandt2002} already discussed in \cite{Pompe2011}. As before, let us consider $\{ \gamma_i \}= \{ \alpha_{i+\Lambda} \}$, with $\Lambda > 0$.
The coupling complexity $C(\gamma,\alpha, \beta)$ can be written as follows (see Eq.~(\ref{c1hd}))
\begin{equation}
\label{ccmi3}
C(\gamma,\beta,\alpha )=\min\{I(\alpha;\tau_{\gamma,\alpha},\tau_{\beta,\alpha}), I(\beta;\tau_{\gamma,\alpha},\tau_{\beta,\alpha})\}.
\end{equation}
Furthermore, Eqs.~(\ref{trv3a}) and (\ref{trv3b}) imply that the entropies $H(\gamma,\alpha,\beta)$, $H(\alpha,\tau_{\gamma,\alpha},\tau_{\beta,\alpha})$, and $H(\beta,\tau_{\gamma,\alpha},\tau_{\beta,\alpha})$ are identical. According to Eq.~(\ref{ccmi3}), the variable leading to the minimum mutual information ($C=0$ in this case) is independent of the joint transcript variable $(\tau_{\gamma,\alpha},\tau_{\beta,\alpha})$. Let us assume that $\min\{H(\alpha),H(\beta)\}=H(\beta)$. Then, the joint entropy of the three symbolic representations can be written as
\begin{equation}
\label{jred}
H(\gamma,\beta,\alpha) = H(\beta)+H(\tau_{\gamma,\alpha},\tau_{\beta,\alpha}).
\end{equation}
We will invoke now the property of {\it monotonicity} of the coupling complexity \cite{Monetti2013}. In fact, one can demonstrate that if $\min\{H(\alpha),H(\beta)\}=H(\beta)$ then $C(\gamma,\alpha, \beta) \geq C(\gamma,\alpha)$, which leads in this case to $C(\gamma,\alpha)=0$. 
Thus, {\it monotonicity} implies the independence of the variables $\alpha$ and $\tau_{\gamma,\alpha}$. 
Similarly to Eq.~(\ref{jred}), the following conditions hold
\begin{eqnarray}
\label{jred1}
H(\gamma,\alpha) &= &H(\alpha)+H(\tau_{\gamma,\alpha})  \nonumber \\
H(\alpha,\beta) &=& H(\beta)+H(\tau_{\beta,\alpha}).
\end{eqnarray}
where Eq.~(\ref{jred1}) follows from the independence of the variables $\beta$ and $\tau_{\beta,\alpha}$.
Using Eqs.~(\ref{jred}) and (\ref{jred1}), Eq.~(\ref{cmi}) becomes
\begin{equation}
\label{conv}
I(\gamma,\beta\mid\alpha)= H(\gamma,\alpha)+H(\beta , \alpha)-H(\gamma ,\beta,\alpha )-H(\alpha) = H(\tau_{\gamma,\alpha})+H(\tau_{\beta,\alpha})-H(\tau_{\gamma,\alpha},\tau_{\beta,\alpha}),
\end{equation}
which implies the equality of Eq.~(\ref{cmi}) and Eq.~(\ref{rr}) and thus dimensional reduction.
In case $\min\{H(\alpha),H(\beta)\}=H(\alpha)$, the property of {\it monotonicity} has a more general implication, i.e. $C(\gamma,\alpha, \beta) \geq C(\gamma,\alpha)$ and $C(\gamma,\alpha, \beta) \geq C(\beta,\alpha)$. Using these conditions, one can analogously derive Eq.~(\ref{conv}). The property of {\it monotonicity} is proved for the multivariate case in \cite{Monetti2013}. 
\begin{figure}[tbp]
\centering
\includegraphics[width=13.cm,angle=0]{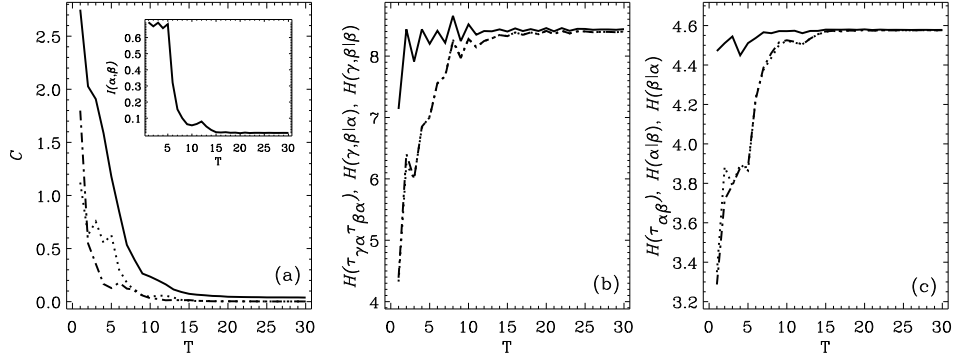}
\caption{
a) The complexity $C$ versus the delay T. The solid curve indicates the complexity $C(\gamma,\beta, \alpha)$, while the dotted curve and the dot-dashed curve display the complexities $C(\gamma,\alpha)$ and $C(\beta,\alpha)$, respectively (more details in text). The inset shows the mutual information $I(\alpha, \beta))$ versus T.
b) The solid curve shows the entropy of transcripts $H(\tau_{\gamma,\alpha},\tau_{\beta,\alpha})$, the dotted curve the conditional entropy $H(\gamma, \beta \mid \alpha)$, and the dot-dashed curve  the conditional entropy $H(\gamma, \beta \mid \beta)$ versus the delay T. The difference between the conditional entropies cannot be observed due to overlapping.
c) The solid curve displays the entropy $H(\tau_{\beta,\alpha})$, the dotted curve the conditional entropy $H(\beta \mid \alpha)$ and the dot-dashed curve the conditional entropy $H(\alpha \mid \beta)$. All results were obtained using $L=4$ and $M=2^{18}$ data points.}
\label{fig:1}
\end{figure}

We have just shown that the coupling complexity is a relevant quantity to take into account when analysing coupling directionality. In the next example, we monitor the behavior of $C$ and other information measures versus the delay time T.  
We consider again the coupled logistic map (\ref{clm}) and generate symbolic representations $\{\alpha_i\}$, $\{\beta_i\}$ for the time series $\{x_i\}$, $\{y_i\}$ and coupling parameters $k_1 = 0.6$ and $k_2 =0.2$. In this example $\{\gamma_i\} = \{\alpha_{i+1}\}$.
Figure~\ref{fig:1} shows the behavior of different information measures as a function of the delay T used to generate ordinal patterns.  Figure~\ref{fig:1}(a) displays the complexity $C(\gamma, \beta,\alpha)$, and the complexities for the pairs $C(\gamma,\alpha)$  and  $C(\beta,\alpha)$, evaluated using Eqs.~(\ref{c1hd}) and (\ref{theorem}), respectively. We observe that the complexity  $C(\gamma, \beta,\alpha)$ is always larger than any of the complexities for the pairs. 
In addition, this plot shows that all complexities approach zero for delay $\text{T} \geq 15$. Thus, requesting $C(\gamma , \beta,\alpha) \sim 0$ for the highest dimension automatically warranties the same condition for lower ones. The inset in Fig.~\ref{fig:1}(a) shows the mutual information of the symbolic representations $I(\beta,\alpha)$ versus the delay T. For this coupled system, $I(\alpha, \beta)$ decreases for increasing T as well.
Figure~\ref{fig:1}(b) and \ref{fig:1}(c) show the behavior of the entropies associated with transcripts and the conditional entropies. For this system, it is hardly possible to distinguish between the conditional entropies. More important, we observe in both plots that for $C(\gamma , \beta,\alpha) \sim 0$, the conditional entropies approach the value of the entropy of the transcripts as predicted by Eqs.~(\ref{c1cond}) and (\ref{jred}).
\begin{figure}[tbp]
\centering
\includegraphics[width=13.cm,angle=0]{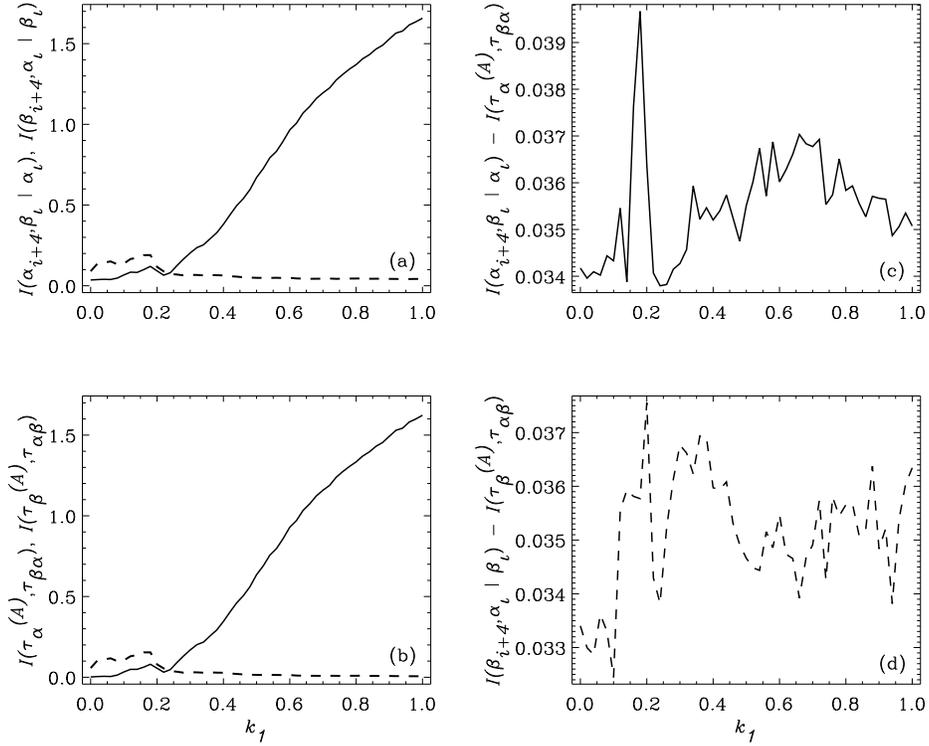}
\caption{
a) Conditional mutual informations $I(\alpha_{i+4},\beta_i \mid \alpha_i)$ (solid curve) and $I(\beta_{i+4},\alpha_i \mid \beta_i)$ (dashed curve) for the coupled logistic map defined in Eq.~(\ref{clm}).
b) The mutual informations $I(\tau_{\alpha}^{(A)} , \tau_{\beta,\alpha})$ (solid curve) and $I(\tau_{\beta}^{(A)} , \tau_{\alpha, \beta})$ (dashed curve), where $\tau_{\alpha}^{(A)} \alpha_i =\alpha_{i+4}$, $\tau_{\beta}^{(A)} \beta_i =\beta_{i+4}$, and $\tau_{\beta,\alpha} \alpha_i =\beta_{i}$.
c) The difference $I(\alpha_{i+4},\beta_i\mid\alpha_i) - I(\tau_{\alpha}^{(A)} , \tau_{\beta,\alpha})$ indicates the error when using Eq.~(\ref{rr}).
d) Idem upper right for $I(\beta_{i+4},\alpha_i\mid\beta_i) - I(\tau_{\beta}^{(A)} , \tau_{\alpha,\beta})$. All results were obtained using $L=4$, $\text{T} = 27$ and times series of length $N = 10^5$ data points.}
\label{fig:2}
\end{figure}

We turn now the focus to the comparison of the two coupling directionality measures (Eqs.~(\ref{cmi}) and (\ref{rr})) within the regime ($C \sim 0$). To this end, we discuss in more detail the coupled logistic map (\ref{clm}) for delay time $\text{T}=27$ (right column in Fig.~\ref{fig:1o}).
%In Fig.~\ref{fig:2}, we present a detailed comparison between the coupling directionality measures for the coupled logistic map (\ref{clm}) for $T=27$ (right column in Fig.~\ref{fig:1o}).
Figure~\ref{fig:2}(a) shows the symbolic transfer entropies for both coupling directions, $x \rightarrow y$ and $y \rightarrow x$, versus the coupling parameter $k_1$. 
For $k_1=0$, there is no information flow $y \rightarrow x$  but a clear response is observed for the information flow in the opposite direction, as expected.  For $k_1 \lesssim 0.3$ the response is non-monotonous for both directions probably due to the dynamical features of this coupled system \cite{Pompe2011}. In particular the crossover point, which is expected to occur around at $k_1 \sim 0.2$ is slightly shifted to higher values. For $k_1 \geq 0.3$, the information flow $y \rightarrow x$ increases monotonically while the information flow $x \rightarrow y$ remains almost constant. It should be remarked that these results can only be compared qualitatively with those presented in reference \cite{Pompe2011}, since the evaluated measures are different.
Figure~\ref{fig:2}(b) shows the mutual information between transcripts as described in the caption. As mentioned above, it is hardly possible to find a difference by eye inspection between the upper left and lower left panels. The difference between conditional mutual information and mutual information of the transcripts (Eqs.~(\ref{cmi}) and (\ref{rr})) is quantified in Figs.~\ref{fig:2}(c) and \ref{fig:2}(d).
The mean and standard deviation of the difference are around 3.5x$10^{-2}$ and 1.1x$10^{-3}$ in both cases. 
\begin{figure}[tbp]
\centering
\includegraphics[width=13.cm,angle=0]{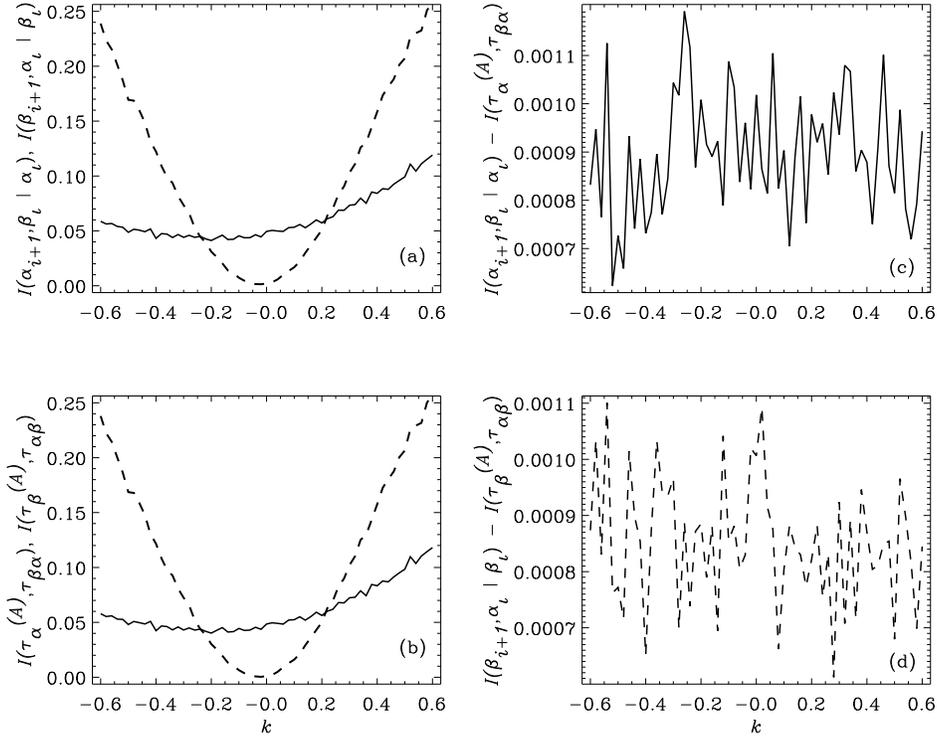}
\caption{
The symbolic representations $\{ \alpha_i \}$ and $\{ \beta_i \}$ correspond to time series $x$ and $y$ for the coupled autoregressive models defined by Eq.~(\ref{ks}). 
a) Conditional mutual informations $I(\alpha_{i+1},\beta_i \mid \alpha_i)$ (solid curve) and $I(\beta_{i+1},\alpha_i \mid \beta_i)$ (dashed curve).
b) The mutual informations $I(\tau_{\alpha}^{(A)} , \tau_{\beta,\alpha})$ (solid curve) and $I(\tau_{\beta}^{(A)} , \tau_{\alpha, \beta})$ (dashed curve), where $\tau_{\alpha}^{(A)} \alpha_i =\alpha_{i+1}$, $\tau_{\beta}^{(A)} \beta_i =\beta_{i+1}$, and $\tau_{\alpha,\beta} \alpha_i =\beta_{i}$.
c) The difference $I(\alpha_{i+1}, \beta_i \mid \alpha_i) - I(\tau_{\alpha}^{(A)} , \tau_{\beta,\alpha})$ indicates the error when using Eq.~(\ref{rr}).
d) Idem upper right for $I(\beta_{i+1},\alpha_i \mid \beta_i) - I(\tau_{\beta}^{(A)} , \tau_{\alpha, \beta})$. All results were obtained using $L=3$, $\text{T} = 30$ and times series of length $N = 10^5$ data points.}
\label{fig:3}
\end{figure}

As a second example, we present two linearly bidirectionally coupled autoregressive models defined by the following expression,
\begin{equation}
x_{i+1}=k_1x_i+k_c y_{i}+\eta^{x}_{i}, \quad  y_{i+1}=k_2y_i+k x_{i}+\eta^{y}_{i},
\label{ks}
\end{equation}
where $k_1 = 0.6$ and $k_2=0.5$, and $\eta^x$ and $\eta^y$ are normal random numbers. The parameters $k_c =0.2$ and $k$ are the couplings between system components, where $k$ is varied in the range $k \in [-0.6,0.6]$. This system was studied analytically using transfer entropy in \cite{Kaiser2002} for the case $k_c =0$. As before, Fig.~\ref{fig:3}(a) shows the CMI for both coupling directions $x \rightarrow y$ and $y \rightarrow x$ versus the coupling parameter $k$. The solid curve indicates that the information flow $y \rightarrow x$ never vanishes. This is expected since $k_c =0.2$  for the whole range of $k$ values. A clear asymmetry is observed between the regions $k > 0$ and $k<0$, since the symmetry of Eq.~(\ref{ks}) is broken for $k_c \neq 0$.
Thus, the CMI $I(\alpha_{i+1},\beta_i \mid \alpha_i) > I(\beta_{i+1},\alpha_i \mid \beta_i)$ for $-0.25 \lesssim k \lesssim 0.20$ and  $I(\alpha_{i+1},\beta_i \mid \alpha_i) < I(\beta_{i+1},\alpha_i \mid \beta_i)$ for $-0.25 \gtrsim k  \gtrsim 0.20$. For $k \sim 0.20$ and $k \sim -0.25$ the values of the CMI are similar, revealing a balanced situation with no preferred coupling direction. It should also be noted that $I(\beta_{i+1},\alpha_i\mid\beta_i)$ vanishes for $k = 0$ since there is no information flow $x \rightarrow y$ for this value of the coupling parameter. 
Figure~\ref{fig:3}(b) shows the mutual information between transcripts, as described in the caption of Fig.~\ref{fig:3}.  Once again, there is a striking similarity between the left panels. 
Figure~\ref{fig:3}(c) and \ref{fig:3}(d) indicate the difference between the two approaches. In both cases, the mean and standard deviation of the difference are around 9x$10^{-4}$ and 1x$10^{-4}$. 
\begin{figure}[tbp]
\centering
\includegraphics[width=13.cm,angle=0]{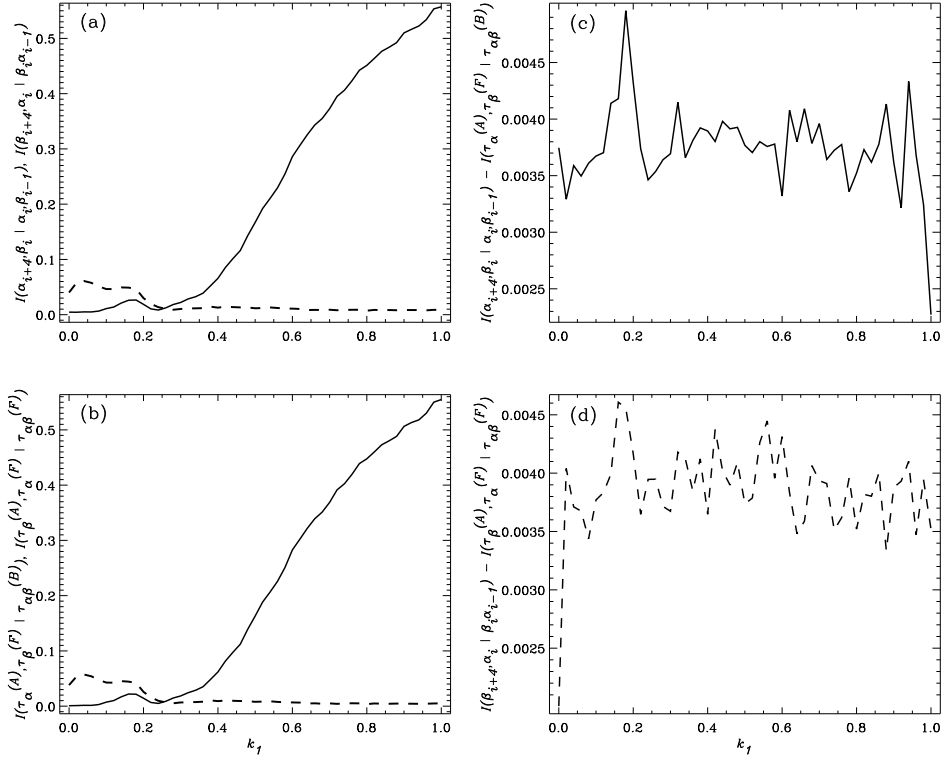}
\caption{
a) Conditional mutual informations $I(\alpha_{i+4},\beta_i\mid\alpha_i,\beta_{i-1})$ (solid curve) and $I(\beta_{i+4},\alpha_i\mid\beta_i,\alpha_{i-1})$ (dashed curve) for the coupled logistic map defined in Eq.~(\ref{clm}).
b) The CMI $I(\tau_{\alpha}^{(A)} , \tau_{\beta}^{(F)}\mid \tau_{\alpha, \beta}^{(B)})$ (solid curve) and $I(\tau_{\beta}^{(A)} , \tau_{\alpha}^{(F)}\mid \tau_{\alpha, \beta}^{(F)})$ (dashed curve), where $\tau_{\alpha}^{(A)} \alpha_i =\alpha_{i+4}$, $\tau_{\beta}^{(A)} \beta_i =\beta_{i+4}$, $\tau_{\alpha}^{(F)} \alpha_{i-1} =\alpha_{i}$, $\tau_{\beta}^{(F)} \beta_{i-1} =\beta_{i}$, $\tau_{\alpha,\beta}^{(B)} \alpha_i =\beta_{i-1}$, and $\tau_{\alpha,\beta}^{(F)} \alpha_{i-1} =\beta_{i}$.
c) The difference $I(\alpha_{i+4},\beta_i\mid\alpha_i,\beta_{i-1}) - I(\tau_{\alpha}^{(A)} , \tau_{\beta}^{(F)}\mid \tau_{\alpha, \beta}^{(B)})$ indicates the error.
d) Idem upper right for $I(\beta_{i+4},\alpha_i\mid\beta_i,\alpha_{i-1}) -  I(\tau_{\beta}^{(A)} , \tau_{\alpha}^{(F)}\mid \tau_{\alpha, \beta}^{(F)})$. All results were obtained using $L=3$, $\text{T} = 30$ and times series of length $N = 10^5$ data points.}
\label{fig:4}
\end{figure}

\subsection{Generalization for more conditions}
We return now to the discussion of Eq.~(\ref{cmi}) and consider first the case where the condition expresses the joint information of two processes, as follows
\begin{equation}
I(\theta,\gamma\mid\alpha,\beta)=H(\theta,\alpha,\beta)+H(\gamma,\alpha,\beta)-H(\theta,\gamma,\alpha,\beta)-H(\alpha,\beta).
\label{cmi4}
\end{equation}
where the CMI has been written in terms of Shannon entropies. Here, we will restrict ourselves to the bivariate case and find the generalized form of Eq.~(\ref{rr}) when accounting for more conditions. For instance, in Eq.~(\ref{cmi4}) we can assume that $\{\theta_i\} = \{\alpha_{i+\Lambda_1}\}$ and $\{\gamma_i\} = \{\beta_{i+\Lambda_2}\}$ with $\Lambda_1 > \Lambda_2 > 0$, thus the variable $(\alpha,\beta)$ accounts for the joint past of the coupled processes.
We use again the condition $C(\theta,\gamma,\alpha,\beta) = 0$, here for four variables, and write as before $C$ in terms of mutual information as follows
%\begin{align}
\begin{equation}
\label{aproxi1}
C(\theta,\gamma,\alpha,\beta) = \min \{I(\alpha; \tau_{\theta,\alpha},\tau_{\gamma,\beta},\tau_{\alpha,\beta}),I(\beta; \tau_{\theta,\alpha},\tau_{\gamma,\beta},\tau_{\alpha,\beta}) \}.
\end{equation}
%\end{align}
In the limit of vanishing coupling complexity, Eq.~(\ref{aproxi1}) implies that the variable associated with the minimum entropy, i.e. $\alpha$ or $\beta$, is independent of the joint transcript variable $(\tau_{\theta,\alpha},\tau_{\gamma,\beta},\tau_{\alpha,\beta})$. 
In this case, one only needs to invoke {\it monotonicity} (see \cite{Monetti2013}) and to follow
the same reasoning which led us to Eqs.~(\ref{jred}) and (\ref{jred1}) to derive
\begin{equation}
I(\theta,\gamma\mid\alpha,\beta) = H(\tau_{\theta,\alpha},\tau_{\alpha,\beta}) +H(\tau_{\gamma,\beta},\tau_{\alpha,\beta})-H(\tau_{\theta,\alpha},\tau_{\gamma,\beta},\tau_{\alpha,\beta}) -H(\tau_{\alpha,\beta}) = I(\tau_{\theta,\alpha},\tau_{\gamma,\beta}\mid\tau_{\alpha,\beta}).
\label{rr4}
\end{equation}
Thus, the CMI for two conditions is reduced to one of three transcripts, where $\tau_{\alpha,\beta}$ accounts for the joint conditional process. Following this strategy, one can easily infer that for $m$ conditions the analysis can be reduced to one of $m-1$ conditions, where only transcripts among symbolic representations are involved. The structure of this approximation scheme naturally induces us to ask for further dimensional reduction. From the point of view of the construction, this is always possible since the scheme does not differentiate between ordinal patterns and transcripts. However, one has to have in mind that every additional dimensional reduction is performed under assumptions different from that expressed by $C  \sim 0$. Thus, it is expected that error increases when reducing the dimensionality of the problem.  However, for some of the considered systems, we have observed that further dimensional reduction still renders very good approximations which describe the main features of the coupling directionality.
\begin{figure}[tbp]
\centering
\includegraphics[width=13.cm,angle=0]{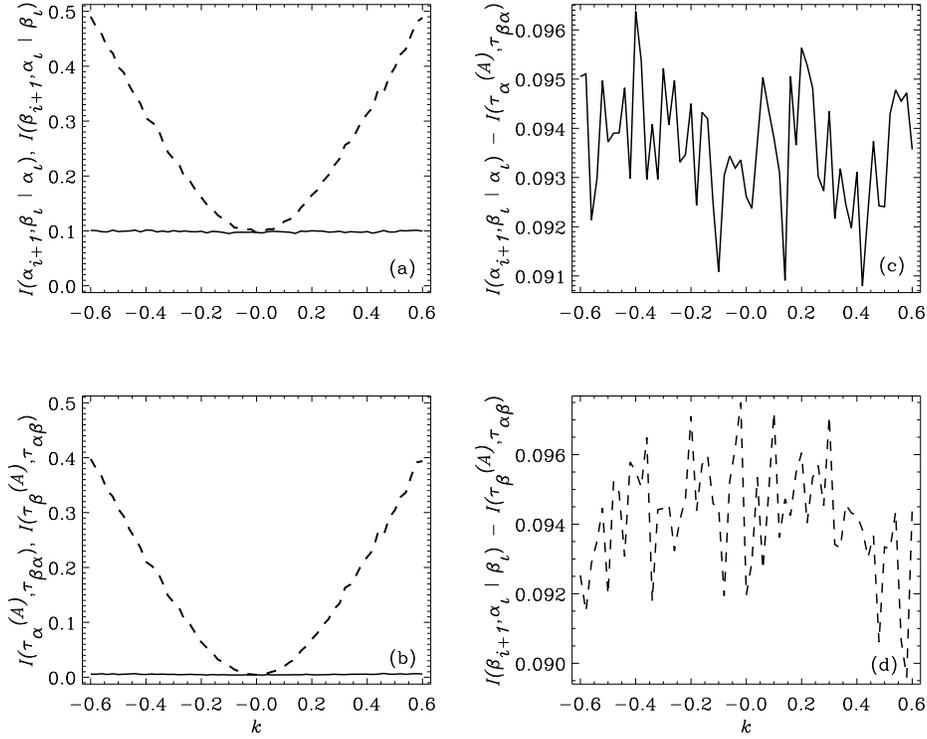}
\caption{
a) Conditional mutual informations $I(\alpha_{i+1},\beta_i\mid\alpha_i)$ (solid curve) and $I(\beta_{i+1},\alpha_i\mid\beta_i)$ (dashed curve) for the two linear coupled autoregressive processes defined in Eq.~(\ref{ks}).
b) The mutual informations $I(\tau_{\alpha}^{(A)} , \tau_{\beta,\alpha})$ (solid curve) and $I(\tau_{\beta}^{(A)} , \tau_{\alpha, \beta})$ (dashed curve), where $\tau_{\alpha}^{(A)} \alpha_i =\alpha_{i+1}$, $\tau_{\beta}^{(A)} \beta_i =\beta_{i+1}$, and $\tau_{\alpha,\beta} \alpha_i =\beta_{i}$.
c) The difference $I(\alpha_{i+1},\beta_i\mid\alpha_i) - I(\tau_{\alpha}^{(A)} , \tau_{\beta,\alpha})$ indicates the error when using Eq.~(\ref{rr}).
d) Idem upper right for $I(\beta_{i+1},\alpha_i\mid\beta_i) - I(\tau_{\beta}^{(A)} , \tau_{\alpha, \beta})$. All results were obtained using $L=4$, $\text{T} = 30$ and times series of length $N = 10^5$ data points.}
\label{fig:5}
\end{figure}

As an example of the application of Eq.~(\ref{rr4}), we consider once again the coupled logistic map (\ref{clm}) already analyzed using Eq.~(\ref{rr}), but we include an additional condition to account for the joint past of the processes. Figure~\ref{fig:4} is similar to Fig.~\ref{fig:2} but the compared measures have the form of those in Eq.~(\ref{rr4}). Figure~ \ref{fig:4}  reveals that including the joint past as condition in the CMI improves the characterization of the coupling directionality displaying a more sensitive response within the range of coupling values where crossover behavior occurs ($k \sim 0.2$). The accuracy of our approach can be observed in Figs.~\ref{fig:4}(c) and \ref{fig:4}(d), with a mean and standard deviation around 4x$10^{-3}$ and  4x$10^{-4}$. 
\subsection{The influence of dimensionality}
A comparison of Eq.~(\ref{rr}) and Eq.~(\ref{rr4}) indicates that the space dimension to estimate information flow increases with the number of conditions. In general, the CMI requires the calculation of the entropy of the $m$-dimensional joint process, where $m$ is the number of symbolic representations involved in the calculation. In addition, the number of available states in this space grows with $L$ as $(L!) ^{m}$. Then, the curse of dimensionality becomes an issue to obtain reliable estimates and one has to find a suitable compromise between $m$, $L$ and the length $N$ of the time series. Since the right hand side of  Eq.~(\ref{rr}) and Eq.~(\ref{rr4}) imply dimensional reduction, they may provide a more accurate quantification of the coupling directionality.

To investigate the influence of dimensionality, we have considered the autoregressive models defined in Eq.~(\ref{ks}) but using $k_c =0$ for the sake of simplicity. 
Figure~\ref{fig:5} shows the same measures as in Fig.~\ref{fig:2} but evaluated for $L=4$ and using the same number of data points. The symbolic transfer entropies (Fig.~\ref{fig:5}(a)) clearly unveils the effect of increasing dimension. In fact, one expects that the information flow $y \rightarrow x$ vanishes in this case. However,  the solid curve, which indicates the information flow $y \rightarrow x$, displays an approximately constant value higher than zero due to poor statistics. On the other hand, our estimate expressed by Eq.~(\ref{rr}) is more robust against increasing dimension, since the dashed curve is still very close to zero mutual information, as observed in Fig.~\ref{fig:5}(b). In this case, the difference between the two coupling directionality measures displayed in Figs.~\ref{fig:5}(c) and \ref{fig:5}(d) is larger because of poor statistics as well. 
\subsection{Other approaches}
Some authors have considered approaches to describe coupling directionality using ordinal patterns, where the information flow is calculated through the sorting information of future values among ordinal patterns describing the history of the systems \cite{Pompe2011,Papana2011}. Some of these information measures even consider the use of ordinal patterns of different lengths $L$. We will show that our approach fits in these constructions and can be implemented in an elegant way. First, we focus on the definition of a transcript between ordinal patterns $\alpha^{L_1}$ and $\alpha^{L_2}$ of lengths $L_1$, and $L_2$, where we assume  $L_1 > L_2$ without loss of generality.  Since $\mathcal{S}_{L_2} \subset \mathcal{S}_{L_1}$ then every element in $\mathcal{S}_{L_2}$ can also be expressed as an element of the larger group $\mathcal{S}_{L_1}$. Let $\Delta L = L_1 - L_2$ be the difference between the length of ordinal patterns. Within $\mathcal{S}_{L_1}$, the symbol  $\alpha^{L_2}$ can be expressed as follows
\begin{equation}
\alpha^{L_2} = \left\langle \alpha^{L_2} _{0}, \cdots ,\alpha^{L_2} _{L_{2}-1}, L_{2}, L_{2}+1, \cdots , L_{2}+\Delta L -1 \right\rangle \nonumber \\
\label{embed}
\end{equation}

By means of this procedure it is always possible to evaluate transcripts between ordinal patterns of different length. Note that the group embedding defined by Eq.~(\ref{embed}) conserves the transcript scheme \cite{Monetti2009} of the smaller group. Let $\{x_t\}$ be a time series and consider the symbol $\alpha_{t_1}^{L}=\left\langle \alpha^{L} _{0}, \cdots ,\alpha^{L} _{L-1}\right\rangle$ which describes the rank ordering of the sequence $(x_{t_1-L+1},x_{t_1-L+2}, \cdots ,x_{t_1})$. The sorting of the value $x_{t_1+\Lambda}$ with $\Lambda > 0$ can be expressed in terms of transcripts using Eq.~(\ref{embed})
\begin{equation}
\tau \circ \left\langle \alpha^{L} _{0}, \cdots ,\alpha^{L} _{L-1},  L \right\rangle = \alpha_{t_{1}}^{L+1},
\label{taud}
\end{equation}
where $\alpha_{t_{1}}^{L+1}$ describes the rank ordering of the sequence $(x_{t_1-L+1},x_{t_1-L+2}, \cdots ,x_{t_1}, x_{t_1+\Lambda})$ (for simplicity we assumed $\text{T} = 1$). Thus, the transcript $\tau$ accounts for  the sorting information of the new value among the sequence of the previous ones.
As an example, we apply these concepts to the {\it momentary sorting information transfer} (MSIT) introduced in \cite{Pompe2011}. This measure was chosen since other approaches considered in the literature are special cases of the MSIT  \cite{Pompe2011}.
Let us consider first the {\it momentary information transfer} defined as \cite{Pompe2011}
\begin{eqnarray}
I_{x \rightarrow y}^{\mbox{MIT}}(\Lambda)=\sum p(x_t,y_{t+\Lambda},z) \log \frac{p(x_t,y_{t+\Lambda}\mid z)}{p(x_t\mid z)p(y_{t+\Lambda}\mid z)} \nonumber \\
\mbox{with the condition} \; \; z=(x_{t-1}^{M_{z_x}},y_{t+\Lambda-1}^{M_{z_y}}).
\label{mit}
\end{eqnarray}
Here $x_t$ and $y_{t+\Lambda}$ are values of the time series, $x_{t-1}^{M_{z_x}}$ and $y_{t+\Lambda-1}^{M_{z_y}}$ are delay vectors of length $M_{z_x}$ and $M_{z_y}$ which determine the joint past $z$ of the dynamical systems. The {\it momentary sorting information transfer} (MSIT) is derived from Eq.~(\ref{mit}) when only accounting for sorting information of $y_{t+\Lambda}$ among $y_{t+\Lambda-1}^{M_{z_y}}$ and $x_t$ among $x_{t-1}^{M_{z_x}}$  \cite{Pompe2011}. This quantity can be written in the form of a CMI as
\begin{equation}
I_{A \rightarrow B}^{\mbox{MSIT}}(\Lambda)= I(\theta_{i+\Lambda},\gamma_i \mid \alpha_{i+\Lambda-1}, \beta_{i-1}),
\label{msit}
\end{equation}
where $\theta_{i+\Lambda}$, $\gamma_i$, $\alpha_{i+\Lambda-1}$ and $\beta_{i-1}$  are the ordinal patterns for  $(y_{i+\Lambda-M_{z_y}+1},y_{i+\Lambda-M_{z_y}+2}, \cdots ,y_{i+\Lambda})$,  \linebreak
$(x_{i-M_{z_x}+1},x_{i-M_{z_x}+2}, \cdots ,x_{i})$, $(y_{i+\Lambda-M_{z_y}},y_{i+\Lambda-M_{z_y}+1}, \cdots ,y_{i+\Lambda-1})$,  and $(x_{i-M_{z_x}},x_{i-M_{z_x}+1}, \cdots ,x_{i-1})$, respectively. Then, it is clear that for $\Lambda > 0$,
$A=x$ and $B=y$, and for $\Lambda < 0$, $A=y$ and $B=x$. 

\begin{figure}[tbp]
\centering
\includegraphics[width=13.cm,angle=0]{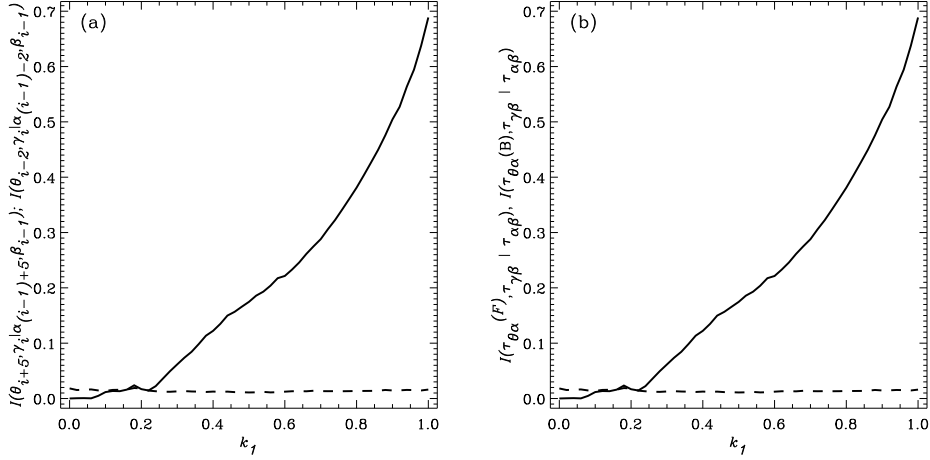}

\caption{a) Conditional mutual informations $I(\theta_{i+5},\gamma_i\mid\alpha_{(i-1)+5},\beta_{i-1})$ (solid curve) and $I(\theta_{i-2},\gamma_i\mid\alpha_{(i-1)+5},\beta_{i-1})$ (dashed curve) for the coupled logistic map defined in Eq.~(\ref{clm}).
b) The CMI $I(\tau_{\theta,\alpha}^{(F)} , \tau_{\gamma,\beta}\mid \tau_{\alpha ,\beta})$ (solid curve) and $I(\tau_{\theta,\alpha}^{(B)} , \tau_{\gamma,\beta}\mid \tau_{\alpha ,\beta})$ (dashed curve), where $\tau_{\theta,\alpha}^{(F)} \alpha_{(i-1)+5} =\theta_{i+5}$, $\tau_{\gamma,\beta} \beta_{i-1} =\gamma_{i}$, $\tau_{\theta,\alpha}^{(B)} \alpha_{(i-1)-2}  =\theta_{i-2}$, and $\tau_{\alpha,\beta} \alpha_{(i-1)+5} =\beta_{i-1}$. The ordinal patterns $\theta$ and $\gamma$ are of length $L_1 = 3$, and $\alpha$ and $\beta$ of length $L_2 = 2$. Thus, all patterns were embedded in $\mathcal{S}_{3}$. We used $\text{T}=30$ and times series of length $N = 10^5$ data points.
}
\label{fig:6}
\end{figure}
We immediately identify that our approach as given in Eq.~(\ref{rr4}) can be applied to the MSIT as follows
\begin{equation}
I_{A \rightarrow B}^{\mbox{MSIT}}(\Lambda) \sim I(\tau_{\theta,\alpha},\tau_{\gamma,\beta}\mid\tau_{\alpha,\beta}),
\label{cmi1}
\end{equation}
where the transcripts $\tau_{\theta,\alpha}$, $\tau_{\gamma,\beta}$ which  provides the sorting information of $x_t$ among $x_{t-1}^{M_{z_x}}$ and $y_{t+\Lambda}$ among $y_{t+\Lambda-1}^{M_{z_y}}$  are evaluated according to Eq.~(\ref{taud}). The transcript $\tau_{\alpha,\beta}$ corresponds to the joint past and is evaluated in the general case using the group embedding defined in Eq.~(\ref{embed}). It should be noted that the approach given by Eq.~(\ref{cmi1}) is not restricted to the use of consecutive values for generating ordinal patterns. In fact, one can always search for a suitable delay T satisfying the condition $C  \sim 0$.
We applied the above described approach to the coupled logistic map (Eq.~(\ref{clm})) using the same coupling parameters as before, for the sake of comparison. 
We have chosen ordinal patterns of length $L_1=3$ and $L_2=2$, thus all ordinal patterns are embedded in $\mathcal{S}_{3}$. 
Since the purpose here is to test the approximation given by Eq.~(\ref{rr4}), a delay time $\text{T} = 30$ has been used to generate ordinal patterns and satisfy the condition of vanishing complexity.
For the joint condition ($\alpha,\beta$), ordinal patterns were generated according to Eq.~(\ref{embed}) and the transcripts according to Eq.~(\ref{taud}). We have considered values of $\Lambda$ in the range $\Lambda \in [-7,7]$ but we show results only for the $\Lambda$ values leading to the maximum response for every direction, namely $\Lambda=5$ and $\Lambda=-2$.
Figure~\ref{fig:6} presents a comparison of the two measures appearing in Eq.~(\ref{cmi1}). The agreement between $I_{A \rightarrow B}^{\mbox{MSIT}}$ and $I(\tau_{\theta,\alpha},\tau_{\gamma,\beta}\mid\tau_{\alpha,\beta})$ is also remarkable for this approach. The mean value of the error calculated over the different values of $k_1$ is around 5 x $10^{-3}$ with a standard deviation of 3 x $10^{-3}$. These results are in perfect agreement with those reported in \cite{Pompe2011}.
\begin{figure}[tbp]
\centering
\includegraphics[width=13.cm,angle=0]{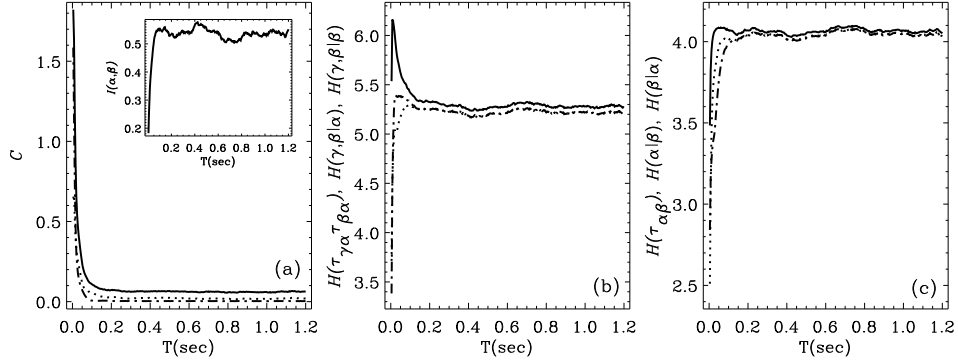}

\caption{a) The complexity $C$ versus the delay T for the frontal electrode pair F4-FP2 in the pre-ictal state. The solid curve indicates $C(\gamma, \beta , \alpha)$, while the dotted curve and the dot-dashed curve display the complexities for the pairs $(\gamma,\alpha)$ and $(\alpha, \beta)$, respectively (more details in text). The inset shows the behavior of the mutual information $I(\alpha,\beta)$ versus T.
b) The solid curve shows the entropy of transcripts $H(\tau_{\gamma,\alpha},\tau_{\beta,\alpha})$, the dotted curve the conditional entropy $H(\gamma, \beta \mid \alpha)$, and the dot-dashed curve  the conditional entropy $H(\gamma, \beta \mid \beta)$ versus the delay T. 
c) The solid curve displays the entropy $H(\tau_{\alpha,\beta})$, the dotted curve the conditional entropy $H(\beta \mid \alpha)$ and the dot-dashed curve the conditional entropy $H(\alpha \mid \beta)$. Results were obtained using $L=4$ and $M \sim 10^{5}$ data points.
}
\label{fig:7}
\end{figure}
\subsection{Application to real world data}
We analyze the electrical brain activity of an infant patient suffering from
frontal lobe epilepsy (FLE). It should be remarked that it is not the purpose of this work to perform a clinical study but to demonstrate the applicability of the above presented methodology to an example of real world data.
A clinical study of the evolution of the brain
electrical activity during therapy has already been presented in Bunk et al. \cite{Bunk}.

The EEG recording was acquired during a time interval of 15 minutes at a sampling rate
of 250 Hz and a signal depth of 16 bits, and consists of 21 synchronously obtained time
series. The positioning of the electrodes
followed that of the standardized 10-20-International System of Electrode
Placements.  We consider an EEG recording which documents a seizure and perform the information directionality assessment for the pre-ictal and ictal states separately. 
\begin{figure}[tbp]
\centering
\includegraphics[width=13.cm,angle=0]{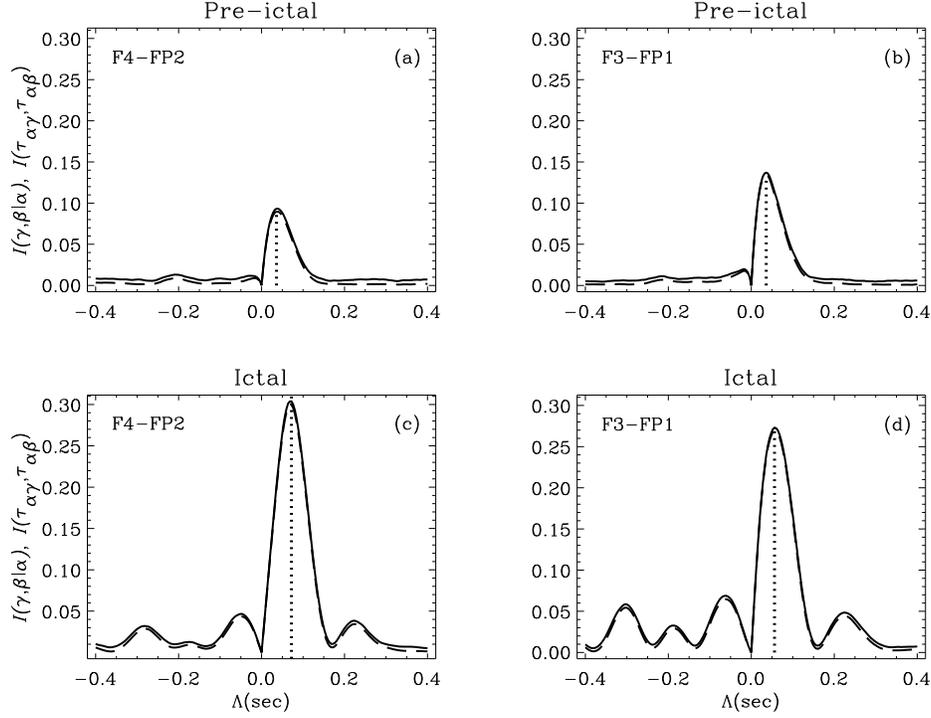}
\caption{ a) The solid curve displays the CMI for the EEG recorded at FP2 and F4. $\{ \alpha_i \}$ and $\{ \beta_i \}$ are the symbolic representations of the time series obtained at F4, FP2, respectively. The symbolic sequence $\{ \gamma_i \} = \{ \alpha_{i+\Lambda} \}$, where $\Lambda \in [-0.4 \sec, 0.4 \sec]$. The dashed curve shows the mutual information of the transcripts $\{(\tau_{\alpha,\gamma})_i \}$ and $\{(\tau_{\alpha,\beta})_i \}$. Both measures were evaluated in the pre-ictal state.
b) Idem a) for  $\{ \alpha_i \}$ and $\{ \beta_i \}$ corresponding to F3, FP1, respectively. Both measures were evaluated in the pre-ictal state.
c) Idem a) but in the ictal state.
d) Idem b) but in the ictal state.
Results were obtained using the parameters $L=3$, $\text{T}=1.2 \sec$ and time series of length $M \sim 10^{5}$ and $M \sim 1.3 \mbox{x} 10^{5}$ data points.
}
\label{fig:8}
\end{figure}

Figure~\ref{fig:7} shows the behavior of some information measures evaluated for the EEG pair F4-FP2 in the pre-ictal state as a function of the delay T
used to generate the symbolic representation. Here  $\{\alpha_i\}$, $\{\beta_i\}$, $\{\gamma_i \}$ are the symbolic representations of the time series $\{x_i\}$ of F4, $\{y_i\}$ of FP2,  and $\{x_{i+1}\}$, respectively.
All measures except the mutual information $I(\alpha,\beta)$ behave as in Fig.~\ref{fig:1}. In fact, $I(\alpha,\beta)$ displays exactly the opposite trend, asymptotically approaching a saturation value greater than zero. It is remarkable that all approximations given in section \ref{sec:theo} are valid even though the $I(\alpha,\beta)$ unveils completely different  interactions. According to Fig.~\ref{fig:7}(a), we generate ordinal patterns using  a T value to satisfy region ($C \sim 0$) and calculate
for every pair of electrodes and for every state the measures appearing in Eq.~(\ref{rr}), where $\{\gamma_i\} = \{\alpha_{i+\Lambda}\}$. These information directionality measures were evaluated for different time lags $\Lambda$, in order to determine the main driving electrodes and the lag of the maximum response. 

Figure~\ref{fig:8} shows the CMI and the mutual information of the transcripts for the EEG pairs FP2-F4 and FP1-F3 in the pre-ictal and ictal states. These EEG pairs were chosen since they lead to the strongest responses.  All plots display a maximum for positive  $\Lambda_{\max}$ values, clearly indicating that FP2 and FP1 are the driving signals. 
We observe that both measures provide almost the same information about the coupling directionality. In particular, both curves indicate the same position for the maximum response $\Lambda_{\max}$. 
Within the covered range of $\Lambda$ values, the error is rather constant ($\sim 4 \mbox{x} 10^{-3}$), except around $\Lambda = 0$ where lower values are observed. 
This shows indirectly the weak dependence of $C$ on $\Lambda$ for this real world data. In all cases,  the mutual information of the transcripts displays lower or equal values than the CMI. 

A global analysis considering all pairs shows that for the pre-ictal (ictal) state 17 (14) out of the 20 strongest responses are driven by frontal signals. This result agrees with the brain pathology of the infant and suggests that signals from the epileptic focus might be driving other brain areas \cite{Bunk}. 
A comparison of Figs.~\ref{fig:8}(a) and \ref{fig:8}(b) with \ref{fig:8}(c) and \ref{fig:8}(d), indicates that  for the ictal state responses increase and $\Lambda_{\max}$ becomes longer. For the pre-ictal state, the mean lag $<\Lambda_{\max} > = 0.041 \pm 0.014 \sec$, while for the ictal state  $<\Lambda_{\max} > = 0.061 \pm 0.017 \sec$, where averages were taken over the 20 strongest responses.

\section{Conclusions}
The concept of transcripts arises naturally when studying relationships between dynamical systems using ordinal symbolic dynamics. Using transcripts one can exploit properties of the symmetric group and combine them with information theoretical approaches. In this work, we have considered the problem of estimating coupling directionality for the bivariate case, and introduced novel information directionality measures which depend only on transcripts for single and joint conditions. Generalizations of these  information directionality measures to the muti-variate case are feasible and will be presented elsewhere. These new directionality measures have the important property of calculating the information flow estimate in lower dimension, which may be preferable for small data sets.  We have also proved that the well established conditional mutual information quantifiers reduced to the proposed measures when a condition of vanishing complexity is fulfilled. A rather general search strategy for low complexity has also been provided. 

Furthermore, we have introduced the concept of group embedding which allows generalizing the definition of transcripts to ordinal patterns of different lengths. Using this extension, different approaches to calculate information flow could be considered within the same framework. We have applied our method to synthetic model data and real world data as well. An example was presented demonstrating the suitability of this transcript based approach to tackle information directionality in EEG data as a diagnostic tool.

\acknowledgements
J.M.A. was financially supported by Ministerio de Ciencia e Innovaci\'{o}n, grant MTM2012-31698.

\bibliography{Monetti_PRE}

\end{document}